\documentclass[final,5p,times,twocolumn]{elsarticle} 

\usepackage{hyperref}
\usepackage{lineno}
\usepackage{multirow}
\usepackage{multicol}
\usepackage{gensymb}
\usepackage{amssymb}
\usepackage{subcaption}
\usepackage{tablefootnote}
\usepackage{caption}
\usepackage{subcaption}
\usepackage{xcolor}

\journal{NIM Section A}

\begin{document}
\begin{frontmatter}

\title{Silicon sensors with resistive read-out:\\ Machine Learning techniques for ultimate spatial resolution}

\author[address1,address3]{M. Tornago\corref{mycorrespondingauthor}}
\cortext[mycorrespondingauthor]{Corresponding author}
\ead{marta.tornago@edu.unito.it}
\author[address5]{F. Giobergia}
\author[address3,address1]{L. Menzio}
\author[address1]{F. Siviero}
\author[address4]{R. Arcidiacono}
\author[address3]{N. Cartiglia}
\author[address1]{M. Costa}
\author[address4]{M. Ferrero}
\author[address1]{G. Gioachin}
\author[address3]{M. Mandurrino}
\author[address3]{V. Sola}

\address[address1]{Università degli Studi di Torino, Torino, Italy}
\address[address3]{INFN, Torino, Italy}
\address[address4]{Università del Piemonte Orientale, Novara, Italy}
\address[address5]{Politecnico di Torino, Torino, Italy}

\begin{abstract}

Resistive AC-coupled Silicon Detectors (RSDs) are based on the Low Gain Avalanche Diode (LGAD) technology, characterized by a continuous gain layer, and by the innovative introduction of resistive read-out. Thanks to a novel electrode design aimed at maximizing signal sharing, RSD2, the second RSD production by Fondazione Bruno Kessler (FBK), achieves a position resolution on the whole pixel surface of about 8 $\mu m$ for 200-$\mu m$ pitch. RSD2 arrays have been tested using a Transient Current Technique setup equipped with a 16-channel digitizer, and results on spatial resolution have been obtained with machine learning algorithms.

\end{abstract}


\end{frontmatter}

%\linenumbers

\section{Introduction}

\begin{figure}[h]
\begin{center}
\includegraphics[width=\linewidth]{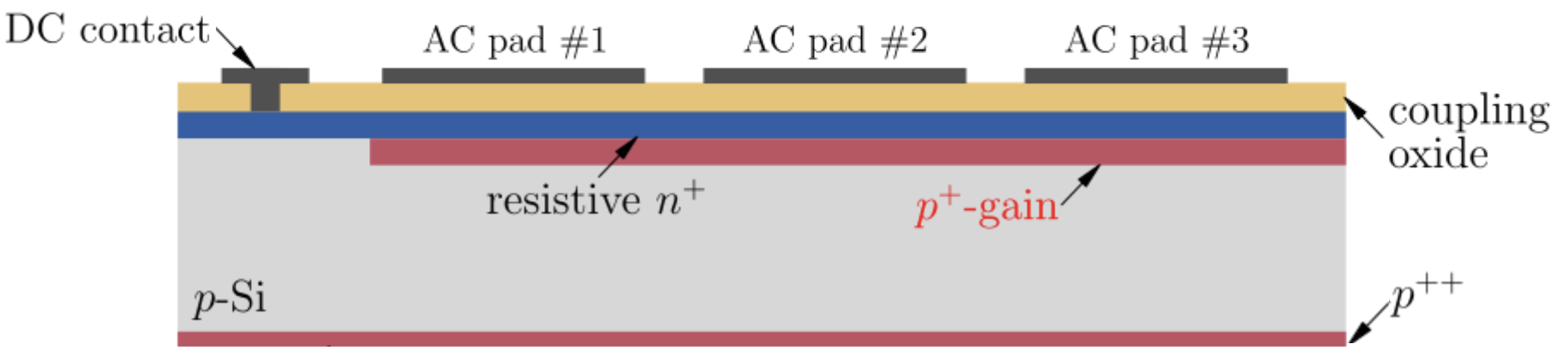} 
\caption{Cross section of an RSD sensor.}
\label{fig:RSD}
\end{center}
\end{figure}

RSDs are a new generation of n-in-p silicon sensors with nearly 100\% fill-factor designed for high-precision 4D tracking in experiments at future colliders. RSDs are based on the Low Gain Avalanche Diode (LGAD) technology but contain one single continuous gain layer. The segmentation of the device is realized by resistive AC-coupled read-out (Fig. \ref{fig:RSD}): (i) the AC coupling of the metal pads occurs through a dielectric layer, and (ii) a continuous resistive n+ electrode allows charge sharing. As a result, the signal is shared among multiple read-out pads. When a particle hits the sensor, each AC pad sees a signal which becomes smaller and more delayed with increasing distance from the impinging point. This RSD key feature allows reaching an unprecedented spatial resolution.

RSD2 was manufactured in 2021: it includes 15 wafers with varying resistivity ($\sim k \Omega / \square$), oxide thickness, and gain dose. Each wafer comprises several sensor geometries, with different active areas, pitch, and AC pad number, and size. With respect to RSD1~\cite{marta}, this new batch optimizes the design of the AC pads in order to minimize the area covered by metal and improve the sharing of signals among pads.

%, as resulting from the measurements performed on RSD1, described in \cite{marta}.

\section{Laboratory measurements}

RSD2 sensors have been tested in the Laboratory for Innovative Silicon Sensors (LISS) in Torino. Three $750\times750\;\mu m^2$ RSD arrays with $200 \mu m$ pitch and $3\times4$ AC pads have been selected for spatial resolution studies. The three matrices differ in the layout of the AC pads, which have shapes of ``Swiss crosses'', ``flakes'' and ``boxes'' (Fig. \ref{fig:spaceres}). Measurements have been performed with the Particulars Transient Current Technique (TCT) setup \cite{TCT}, which exploits a laser to simulate the passage of a minimum ionizing particle (MIP) through the device under test (DUT). This setup is provided with (i) a picosecond\footnote{referred to the laser pulse width} infrared laser with 1064 nm wavelength, (ii) an optical system that can reach a minimum laser spot of $\sim 10\; \mu m$, and (iii) an x-y moving stage with $1\; \mu m$ precision where the sensor is mounted. Each array is wire-bonded to a 16-channel readout board designed at Fermilab \cite{FnalBoard}. Data are acquired with a 16-channel CAEN DT5742 Desktop Digitizer, simultaneously recording all the detector channels.
The whole DUT surface is scanned with the TCT setup: the laser pulses struck every $10\; \mu m$ along \textit{x} and \textit{y} and 100 waveforms are acquired for each AC pad; a typical RSD pulse is shown in~\cite{marta}. When impinging on the sensor surface, the laser provides a signal equivalent to $\sim\; 5$ MIPs. The scan is repeated at 250 V, 300 V, and 330 V, corresponding to gain values of $\sim$ 10, 15, and 20, respectively.

\section{Machine Learning Analysis}

Position reconstruction is based on the combination of information on signals from each AC pad. The correct analytic law describing the relation between waveforms properties and predicted coordinates is not easy to define \cite{marta}. This task is instead perfectly suited for a Machine Learning (ML) algorithm \cite{fede_ml}: signal properties are fed as input features, while the predicted \textit{x-y} coordinates are the output. The ML analysis of RSD data is based on the following steps: (i) \textit{feature extraction}: meaningful input features are extracted from the experimental data, such as the signal amplitudes; (ii) \textit{train/test split}: the data extracted is split into a training set - used to build a regression model - and a test set, used to assess the performance of the model itself in terms of spatial resolution. We adopted an 80/20 train/test split. For a fair estimate of the model performance, we split the dataset so that all 100 waveforms collected for a specific \textit{x-y} position are all either used for the training or the test; (iii) \textit{model training}: a random forest regression model \cite{breiman2001random} is trained using the training dataset. The random forest is comprised of 100 trees independently trained on random subsets of the training set; (iv) \textit{model evaluation}: the model performance is assessed on the test dataset to obtain the final results. The positions used for the test set differ from the ones used during training to assess the capability of the ML model to generalize to new, unseen positions.

\section{Experimental results}

The spatial resolution for the DUTs has been computed by comparing the \textit{x-y} predicted positions with the laser reference ones, provided by the TCT stage. The differences between predicted and reference coordinates result in a gaussian distribution: its standard deviation represents the spatial resolution of the whole system, accounting for both the RSD and the laser resolution. As the distributions are created separately for \textit{x} and \textit{y} coordinates, the total spatial resolution for a RSD array is the combination of the two resolutions:

\begin{equation}
    \sigma_{sys} = \sqrt{\sigma_{RSD}^2+\sigma_{laser}^2},\;\; \sigma_{RSD} = \sqrt{\sigma_{RSD,x}^2+\sigma_{RSD,y}^2}.
\end{equation}

\begin{figure}[h]
\begin{center}
\includegraphics[width=\linewidth]{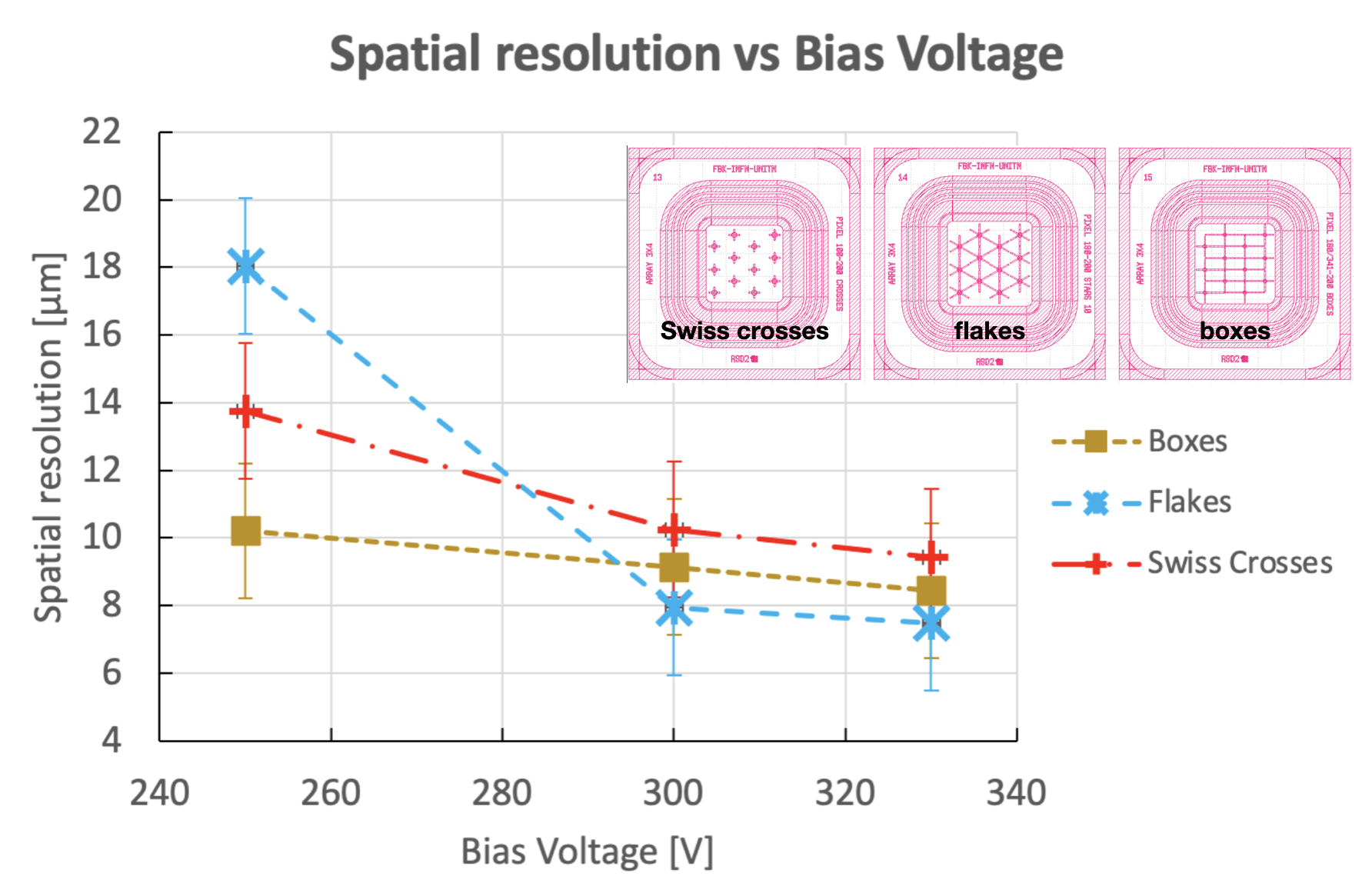} 
\caption{Results for total spatial resolution represented as a function of bias voltage for ``boxes'', ``flakes'' and ``Swiss crosses'' $200\; \mu m$ pitch RSD2 arrays.}
\label{fig:spaceres}
\end{center}
\end{figure}

For the sensors symmetry, $\sigma_{RSD,x} \simeq \sigma_{RSD,y}$ , so  $\sigma_{RSD}\simeq \sqrt{2}\sigma_{RSD,i}$ (where \textit{i} is \textit{x} or \textit{y}). In fig. \ref{fig:spaceres} spatial resolution values are represented as a function of bias voltage for the three geometries. $200-\mu m$-pitch RSD2 matrices can reach a total spatial resolution $\sigma_{RSD,tot} \sim 8\mu m$ at a gain $\sim 20$. This result is much smaller than the corresponding binary readout precision, which would be $pitch\;size/\sqrt{12} \sim 58 \mu m$. Resolution improves with increasing bias voltage (and gain) thanks to larger signals (better signal-to-noise ratio), and plateaus after 300 V (gain $\sim$15). These measurements do not allow to claim whether one of the three structures has better performances, as their resolutions are compatible within the errors, which are mainly represented by the uncertainty on $\sigma_{laser}\sim 2\;\mu m$. The contribution to uncertainty from ML reconstruction has been calculated and can be considered negligible \cite{fede_ml}.
Better spatial resolution results are expected using point-like particles instead of a $10-\mu m$ spot laser and exploiting a setup provided with a precise tracking system.

\section{Conclusions}

This contribution describes the latest studies on the spatial resolution of three arrays from the FBK RSD2 production tested with a TCT setup equipped with a 16-channel digitizer. 
The characteristic charge sharing of RSDs allows performing position reconstruction with the use of Machine Learning. 
Results demonstrate that RSD2 200-$\mu m$-pitch matrices can achieve a spatial resolution of $\sim 8\;\mu m$ at gain $\sim 20$ with a laser intensity corresponding to $\sim 5$ MIPs. Since particles detected in high energy experiments are usually MIPs, further studies are ongoing to evaluate the RSD2 spatial and timing resolutions with 1 MIP both with laser measurements and at test beams.

\section{Acknowledgements}
We gratefully acknowledge the following funding agencies and collaborations: INFN–CSN5, RSD Project;  FBK-INFN collaboration framework; MIUR PRIN project 4DInSiDe; Dipartimenti di Eccellenza, Torino University (ex L.232/2016, art. 1, cc. 314, 337); SmartData@PoliTo.

\end{document}